# Partially Molten Plumes and Melt-Fingers: Two Modes of Magma-Transport through the Mantle in Terrestrial Bodies

Author #1: Ken'yo U[1,2], (kenyo.u.grc@gmail.com ; "U" is the family name)

Author #2: Masanori Kameyama[1,3],

Author #3: Takehiro Miyagoshi[3],

Author #4: Takatoshi Yanagisawa[3],

Author #5: Masaki Ogawa[2],

1. Geodynamics Research Center (GRC), Premier Institute for Advanced Studies (PIAS), Ehime University, Matsuyama, Ehime 790-8577, Japan.

2. Department of Earth Sciences and Astronomy, The University of Tokyo at Komaba, Meguro, Tokyo 153-8902, Japan.

3. Japan Agency for Marine-Earth Science and Technology (JAMSTEC), Yokohama, Kanagawa 236-0001, Japan.

# Abstract
To understand the dynamics of partially molten mantle in terrestrial bodies, we carried out a linear perturbation analysis and two-dimensional numerical simulations of a flow of magma and matrix in a horizontal layer, where magma generated by decompression melting of matrix percolates through the convecting matrix. Our study shows that the upward migration of magma takes place in two regimes, depending on the competition between the two kinds of flows both of which are driven by the melt-buoyancy: the Stokes velocity of matrix and the percolation velocity of melt. When the melt-buoyancy parameter $B_m$, defined by the ratio of the former velocity to the latter, is large, the magmatism-mantle upwelling (MMUb) feedback dominates the convective flow in the layer: the buoyancy of the magma generated by an upwelling of matrix due to decompression melting boosts the upwelling flow itself. When a solid layer is overlaid on the partially molten layer, the MMUb feedback induces partially molten plumes that ascend through the solid layer by their melt-buoyancy. At lower $B_m$, in contrast, a perturbation in the melt-content in the partially molten layer propagates upward as a porosity wave: the perturbation induces a spatial variation in the rate of expansion or contraction of matrix that are caused by upward migration of magma, leading to an upward shift of the perturbation. When a solid layer is overlaid, the porosity wave causes an instability along the layer boundary, which generates a finger-like structure of magma, or melt-finger, that extends upward into the solid layer. The calculated threshold in $B_m$ for the onset of the MMUb feedback suggests that the feedback is responsible for the volcanism that generated the Large Igneous Provinces while being unimportant for hotspot volcanism on the Earth. Since $B_m$ increases with decreasing matrix viscosity, volcanism caused by the MMUb feedback is likely to have been more important in earlier terrestrial planets where the mantle was hotter and softer. Melt-fingers are, in contrast, expected to have developed in the lunar mantle if a partially molten layer has developed at its base in the history of the Moon.

# Main Text

## 1 Introduction

Melting history of the mantle is an important part of the thermal and compositional evolution of the interior of terrestrial bodies (e.g., Davies, 2011; Breuer & Moore, 2015). Several earlier numerical studies have emphasized that heat transport by magma strongly influences the thermal history of planets (e.g., Moore and Webb, 2013; Lourenço et al., 2018, 2020). On the spatial scale of mantle convection, hundreds of kilometers, magma generated in the mantle can also influence its dynamic behavior: mantle upwelling generates magma by decompression melting, while the buoyancy of the generated magma further enhances the upwelling flow (Ogawa, 2018, 2020). This positive feedback, known as the magmatism–mantle upwelling feedback through melt buoyancy (MMUb) or the buoyant melting instability (Tackley and Stevenson, 1993; Jha et al., 1994; Hernlund et al., 2008a,b; Ogawa, 2018, 2020), plays a crucial role in some earlier numerical models of mantle evolution in Mars, Venus, and the Earth (e.g., Ogawa, 2003, 2014, 2021; Ogawa & Yanagisawa, 2011, 2012). In addition to the MMUb feedback, a different type of instability, termed the melt-finger, has been identified in the models of lunar mantle evolution (U et al., 2023, 2025). The melt-fingers occur as finger-like features of magma-bearing mantle materials that grow from a partially molten layer that develops at the base of the mantle, as shown in Figure 1: magma is generated at the base of the mantle owing to strong internal heating at 0.36 Gyr, and isolated patches of partially molten materials, or "melt-fingers," grow from the top of the partially molten region at 0.44 Gyr. The melt-fingers then become thicker and evolve into partially molten plumes that ascend by the buoyancy of the melt they bear to generate more magma at 0.52 Gyr; the plume activity is an example of the MMUb feedback. To more clearly show the nature of the MMUb feedback and melt-fingers and, especially, to clarify the condition under which these features of magma-ascent emerge, here, we present a linear perturbation analysis and two-dimensional (2-D) numerical models of magmatism and mantle convection.

In his review paper, McKenzie (1984) developed a mathematical formulation of the problem of magma generation and migration in the mantle. On the spatial scales of several hundreds of kilometers relevant to mantle convection, this formulation has been first applied to the migration of magma generated by a plate-driven mantle flow beneath subduction zones and mid-ocean ridges (e.g., Spiegelman and McKenzie, 1987). Later studies further considered the buoyancy of magma in their studies of magma-generation beneath mid-ocean ridges and noticed that the buoyancy enhances the upwellings there; this effect is called the buoyant melting instability (Jha et al., 1994; Scott and Stevenson, 1989; Tackley & Stevenson, 1993; Hernlund et al., 2008a, b). More recent studies have further extended these studies to analyze magma generation and migration coupled with brittle lithospheric deformation (Keller et al., 2013) and melt–solid separation during magma ocean solidification (Boukare and Ricard, 2017). However, a feature like the melt-fingers in Figure 1 has not been noticed, and the condition that the buoyant melting instability, or the MMUb feedback, and melt-fingers occur under more general context has remained unclear.

The formulation by McKenzie (1984) has also been used in studies of magma-migration on a smaller spatial scale, and some features that resemble melt-fingers have been reported. Several researchers found that channels of magma develop in a partially molten region by the reaction infiltration instability (e.g., Aharonov and Spiegelman, 1997; Spiegelman et al., 2001; Keller and Katz, 2016) or through the melt-content dependence of mantle viscosity (Stevenson, 1989). Batches of melt also develop as solitary waves (Scott & Stevenson, 1986; Spiegelman, 1993a, 1993b). In all of these studies, however, the instability that induces a heterogeneous distribution of magma is caused by matrix compaction and is significant only at spatial scales of the so-called compaction length (McKenzie, 1984), around 1 km or less; compaction is not important on the larger spatial scale of mantle (Ribe, 1985). In their studies of the migration of magma generated by a given localized melt source, in contrast, Richter and Daly (1989) demonstrated that the magma generated at the source propagates upward through the matrix as solitary waves on the spatial scale of several tens of kilometers; their solitary waves are more like the melt-fingers observed in Figure 1.

To clarify the nature of the MMUb feedback, melt-fingers, and the solitary waves noticed earlier together with the conditions that these features arise on the spatial scale much larger than the compaction length under a general context, here, we self-consistently calculate both magma-generation by decompression melting of matrix and the flow of magma and matrix in a 2-D horizontal layer of partially molten materials. In the numerical simulation, we used the finite difference code developed in our previous works

(Miyagoshi et al., 2018; Kameyama, 2022, 2025; Taito et al., 2025), which calculates the continuity, energy, and momentum equations on a uniform mesh of 128 grid points ($x$: lateral direction) by 64 grid points ($z$: vertical direction).

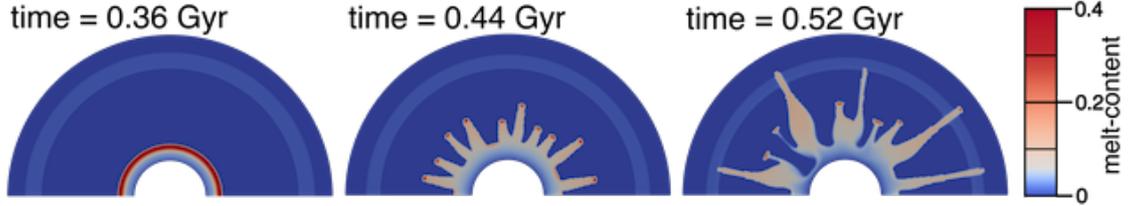

**Figure 1.** Snapshots of the melt-content taken from a model of the lunar mantle evolution in U et al. (2023). Melt-fingers grow from the top of the partially molten layer at the base of the mantle at 0.44 Gyr. Then, the fingers become thicker and ascend as partially molten plumes driven by the buoyancy of melt at 0.52 Gyr.

## 2 Model Description

We carried out the linear perturbation analysis and numerical simulation for single-component partially molten materials under the Boussinesq approximation. The convection occurs in a 2-D horizontal layer in the linear analysis and in a rectangular box of the aspect ratio $\varpi = 2$ in the simulation. The temperature in the analyzed region is equal to the melting temperature that increases linearly with depth, the viscosity of the matrix is constant, and internal heating is neglected. Magma migrates upward through the coexisting matrix as a permeable flow driven by the buoyancy of magma. The density of the materials depends only on the melt-content.

### 2.1 The Basic Equations

The continuity equation is

$$\nabla \cdot [\phi \mathbf{u} + (1 - \phi)\mathbf{U}] = 0, \qquad (1)$$

where $\phi$ is the melt-content, $\mathbf{u}$ and $\mathbf{U}$ are the velocities of the melt and matrix, respectively.

The momentum equation is

$$-\nabla p + \phi \Delta \rho g \mathbf{e_z} + \eta \nabla^2 \mathbf{U} = 0. \qquad (2)$$

Here, $p$ is the dynamic pressure, $g$ is the gravity, $\mathbf{e_z}$ is the unit vector in the vertical direction positive upward, and $\eta$ is the viscosity. We assume that $\eta$ is constant and neglect its dependence on the temperature and melt-content (e.g., Mei et al., 2002; Kohlstedt & Hansen, 2015) for simplicity. $\Delta \rho = \rho_s - \rho_l$ is the density difference between the solid phase (matrix) $\rho_s$ and the melt phase $\rho_l$. We neglected the temperature-dependence of density and assumed that convection is driven only by the buoyancy of melt, because the temperature is equal to the melting temperature that depends only on the depth and is laterally uniform in the entire calculated region: $\rho_s$ is fixed at 3300 kg m$^{-3}$ while $\rho_l$ is

$$\rho_l = \rho_s \left(1 - \frac{\Delta V}{V_0}\right). \qquad (3)$$

Here, $V_0$ is the reference molar volume of the convecting materials, and $\Delta V$ is the difference in molar volume between the solid and melt phases. That is, $\Delta V/V_0$ stands for the amount of density reduction by melting and is assumed to be constant (see Table 1).

Magma generated by decompression melting migrates upward as a permeable flow (e.g. McKenzie, 1984). The relative velocity between melt and matrix $\mathbf{u} - \mathbf{U}$ is calculated from

$$\phi(\mathbf{u} - \mathbf{U}) = \frac{k_\phi}{\mu} \Delta \rho g \mathbf{e_z} = \frac{k_{\phi_0}}{\mu} \left(\frac{\phi}{\phi_0}\right)^3 \Delta \rho g \mathbf{e_z}, \qquad (4)$$

where $k_\phi$ is the permeability that is proportional to $\phi^3$ (e.g., McKenzie, 1984; Miller et al., 2014), $k_{\phi_0}$ is the reference permeability; $\mu$ is the melt viscosity; $\phi_0 = 0.05$ is the reference melt-content. We neglect the contribution from deformation of matrix, in particular from its compaction in Eq. (4) because the spatial scale considered here is much larger than the "compaction length" $\approx \sqrt{(k_\phi \eta)/\mu} \approx 1$ km (Ribe, 1985).

Following U et al. (2023), the energy equation is written as

$$\frac{\partial h}{\partial t} + \nabla \cdot [\phi h_l \mathbf{u} + (1-\phi) h_s \mathbf{U}] = -g\phi \frac{\Delta V}{V_0} u_z + \kappa \nabla^2 h. \tag{5}$$

Here, the 'reduced' enthalpy is given by

$$h = C_p T + \phi \Delta h (1+G), \tag{6}$$

where $C_p$ is the specific heat; $\Delta h$ is the latent heat of melting; $G$ is a function of depth $d - z$ where $d$ is the height of the convecting region:

$$G = \frac{g}{\Delta h} \frac{\Delta V}{V_0} (d - z). \tag{7}$$

The reduced enthalpy of melt $h_l$ and that of solid $h_s$ in Eq. (5) are given by $h_l = h(\phi = 1)$ and $h_s = h(\phi = 0)$, and the first term on the right-hand-side of Eq. (5) is the '$VdP$' term; we added a diffusion term as in U et al. (2023). We assumed that the diffusivity $\kappa$ is constant. (The diffusion term is necessary to make the numerical calculation of Eqs. (1)-(5) feasible, as will be seen in Section 3 below.) Note that magma migration is calculated from Eq. (5) as a transport of latent heat of melting by migrating magma (Katz, 2008). For more details of the basic equations, see U et al. (2022, 2023).

Since the entire convecting region is assumed to be partially molten, the temperature $T$ is equal to the melting temperature $T_m$ that depends linearly on the depth as

$$T_m = T_0 (1+G), \tag{8}$$

where $T_0$ = 1360 K is the melting temperature at the surface (Katz et al., 2003).

## 2.2 Normalization and Parameter Values

The basic equations are converted into their non-dimensional forms using the length scale $d$, the temperature scale $T_{scale} = \Delta h / C_p$, and the time scale $\tau_{scale} = d/w_0$ where $w_0 = k_{\phi 0} \Delta \rho g / \mu$ is a reference velocity called percolation velocity (Spiegelman, 1993a, b).

The non-dimensional continuity equation is

$$\nabla^* \cdot [\phi \mathbf{u}^* + (1-\phi)\mathbf{U}^*] = 0, \tag{9}$$

where the asterisks stand for normalized quantities. The non-dimensional relative velocity $\mathbf{u}^* - \mathbf{U}^*$ is

$$\phi(\mathbf{u}^* - \mathbf{U}^*) = \left(\frac{\phi}{\phi_0}\right)^3 \mathbf{e_z}. \tag{10}$$

The non-dimensional momentum equation is

$$-\nabla^* p^* + B_m \phi \, \mathbf{e_z} + \nabla^{*2} \mathbf{U}^* = 0, \tag{11}$$

where $B_m$ is the melt buoyancy number which is the ratio of the Stokes velocity of matrix ($\Delta \rho g d^2/\eta$) to the percolation velocity of melt ($w_0$):

$$B_m = \frac{\mu d^2}{\eta k_{\phi_0}}. \tag{12}$$

The non-dimensional energy equation is

$$\frac{\partial h^*}{\partial t^*} + \nabla^* \cdot [\phi h_l^* \mathbf{u}^* + (1-\phi)h_s^* \mathbf{U}^*] = -S_d \phi u_z^* + \kappa_n \nabla^{*2} h^*, \tag{13}$$

where $h^* = T^* + \phi(1+G)$ is the normalized enthalpy,

$$S_d = \frac{g d}{\Delta h} \frac{\Delta V}{V_0} \tag{14}$$

is a measure of the depth-dependence of the melting temperature $T_m^*$

$$T_m^* = T_0^*[1 + S_d(1 - z^*)]. \tag{15}$$

The normalized diffusivity $\kappa_n$ in Eq. (13) is

$$\kappa_n = \kappa/(w_0 d). \tag{16}$$

Table 1. The meanings of the symbols and their typical values in planets.

| Symbol | Meaning | Value |
|---|---|---|
| $d$ | Thickness of the partially molten layer | 100 to 3000 km |
| $\rho_s$ | Reference density | 3300 kg m$^{-3}$ |
| $\Delta V/V_0$ | The amount of density reduction by melting | 0.05 to 0.2 |
| $g$ | Gravitational acceleration | 1 to 10 m s$^{-2}$ |
| $\Delta h$ | Latent heat of melting | 657 kJ Kg$^{-1}$ |
| $C_p$ | Specific heat | 1240 J K$^{-1}$ kg$^{-1}$ |
| $\eta$ | Viscosity of matrix | $10^{20}$ to $10^{22}$ Pa s |
| $\mu$ | Melt viscosity | 1 to 20 Pa s |
| $\phi_0$ | Reference melt-content | 0.05 |
| $k_{\phi_0}$ | Reference permeability at $\phi = \phi_0$ | $10^{-14}$ to $10^{-13}$ m$^2$ |

## 3 Linear Perturbation Analysis

We first carried out a linear perturbation analysis of a steady solution to Eqs. (9) to (16) in a partially molten horizontal layer that indefinitely extends in the horizontal direction, as has been done in Spiegelman (1993a). At the top and bottom boundaries, we impose a cyclic boundary condition:

$$\phi_{top} = \phi_{bottom}, \boldsymbol{u}_{top} = \boldsymbol{u}_{bottom}, \text{ and } \boldsymbol{U}_{top} = \boldsymbol{U}_{bottom.} \qquad (17)$$

(The asterisks that stand for non-dimensional quantities are omitted in this section.) Although the condition in Eq. (17) is formally inconsistent with the assumption of the depth dependence of melting temperature, this discrepancy can be safely ignored in the present analysis by assuming $S_d \ll 1$ (see below). We will show that there are two modes of instability that exert control over the style of magma-migration by this linear analysis; we will discuss how these modes grow in the non-linear regime and how they interact with each other in the numerical simulations presented in the next section.

Since the entire layer is partially molten, temperature $T$ is fixed at the melting temperature $T_m$, and the non-dimensional energy equation becomes

$$(1+G)\frac{\partial \phi}{\partial t} - T_0 S_d [\phi u_z + (1-\phi)U_z] + (1+G)\nabla \cdot (\phi \boldsymbol{u}) = \kappa_n \nabla^2[(1+G)\phi]. \quad (18)$$

A steady solution to Eqs. (9), (10) and (18) is

$$\phi = \bar{\phi} = const. \quad (19)$$

$$\bar{U}_z = -\left(\frac{\bar{\phi}}{\phi_0}\right)^3, \quad (20)$$

$$\bar{u}_z = \frac{\bar{\phi}^2}{\phi_0^3}(1-\bar{\phi}). \quad (21)$$

We then decompose $\phi$, $\boldsymbol{U}$, and $\boldsymbol{u}$ into their steady parts and perturbations:

$$\phi = \bar{\phi} + \delta\phi, \quad (22)$$
$$\boldsymbol{U} = \bar{\boldsymbol{U}} + \delta\boldsymbol{U}, \quad (23)$$
$$\boldsymbol{u} = \bar{\boldsymbol{u}} + \delta\boldsymbol{u}. \quad (24)$$

From Eqs. (9) to (11) and (18), we obtain the following linearized equations:

$$(1+G)\frac{\partial \delta\phi}{\partial t} - T_0 S_d \left(\delta U_z + \frac{3\bar{\phi}^2}{\phi_0^3}\delta\phi\right) + (1+G)\nabla \cdot \delta(\phi\boldsymbol{u}) = \kappa_n \nabla^2 \delta\phi, \quad (25)$$

$$-\nabla \delta p + B_m \delta\phi \boldsymbol{e}_z + \nabla^2 \delta\boldsymbol{U} = 0, \quad (26)$$

$$\nabla \cdot \delta\boldsymbol{U} = -\frac{3\bar{\phi}^2}{\phi_0^3}\frac{\partial \delta\phi}{\partial z}, \quad (27)$$

and

$$\delta\boldsymbol{u} = \delta\boldsymbol{U} + \frac{2\bar{\phi}}{\phi_0^3}\delta\phi\, \boldsymbol{e}_z. \quad (28)$$

(Here, we neglect the $z$-dependence of $G$. This approximation holds when $S_d \ll 1$. This condition may not be satisfied in some deep layers like the deep magma ocean. The dependence is, however, mild even in such cases and is unlikely to affect the overall features of the analysis developed below.) From Eqs. (20), (21), (27), and (28), the third term on the left-hand side of the energy equation (Eq. (25)) can be rewritten as

$$\nabla \cdot \delta(\phi\boldsymbol{u}) = \left(\frac{3\bar{\phi}^2}{\phi_0^3} - \frac{4\bar{\phi}^3}{\phi_0^3}\right)\frac{\partial \delta\phi}{\partial z}, \quad (29)$$

and we obtain

$$\frac{\partial \delta\phi}{\partial t} + \left(\frac{3\bar{\phi}^2}{\phi_0^3} - \frac{4\bar{\phi}^3}{\phi_0^3}\right)\frac{\partial \delta\phi}{\partial z} = \frac{T_0 S_d}{1+G}\left(\delta U_z + \frac{3\bar{\phi}^2}{\phi_0^3}\delta\phi\right) + \kappa_n \nabla^2 \delta\phi. \quad (30)$$

To eliminate the matrix velocity from Eq. (30), we will solve for $\delta\boldsymbol{U}$ for given $\delta\phi$. We then decompose $\delta\boldsymbol{U}$ into the potential part and the solenoidal part:

$$\delta U_x = -\frac{\partial f}{\partial x} + \frac{\partial \psi}{\partial z} \tag{31}$$

and

$$\delta U_z = -\frac{\partial f}{\partial z} - \frac{\partial \psi}{\partial x}, \tag{32}$$

where $f$ is the potential function and $\psi$ is the stream function. Substituting Eqs. (31) and (32) into the momentum equation (Eq. (26)) and the continuity equation (Eq. (27)), we obtain

$$\nabla^4 \psi = B_m \frac{\partial \delta \phi}{\partial x}, \tag{33}$$

$$\nabla^2 f = \frac{3\bar{\phi}^2}{\phi_0^3} \frac{\partial \delta \phi}{\partial z}. \tag{34}$$

Then, we look for a solution in the form of

$$\delta \phi = A(t) exp(ilz) \cos(kx). \tag{35}$$

From Eqs. (33) to (35), we obtain

$$\psi = -\frac{k B_m}{(k^2 + l^2)^2} A(t) exp(ilz) \sin(kx) \tag{36}$$

and

$$f = -\frac{3il}{(k^2 + l^2)} \frac{\bar{\phi}^2}{\phi_0^3} A(t) exp(ilz) \cos(kx). \tag{37}$$

Thus, the vertical velocity $\delta U_z$ becomes

$$\delta U_z = \left[ \frac{k^2 B_m}{(k^2 + l^2)^2} - \frac{3l^2}{k^2 + l^2} \frac{\bar{\phi}^2}{\phi_0^3} \right] \delta \phi. \tag{38}$$

Finally, from the energy equation Eq. (30), we obtain

$$\frac{\partial \delta \phi}{\partial t} + V_z \frac{\partial \delta \phi}{\partial z} = \frac{1}{\tau} \delta \phi, \tag{39}$$

with

$$V_z = \frac{3\bar{\phi}^2}{\phi_0^3} - \frac{4\bar{\phi}^3}{\phi_0^3}. \tag{40}$$

and the growth rate of the perturbation $\tau^{-1}$

$$\tau^{-1} \equiv \frac{T_0 S_d}{1 + G} \frac{k^2}{k^2 + l^2} \left( \frac{B_m}{k^2 + l^2} + \frac{3\bar{\phi}^2}{\phi_0^3} \right) - \kappa_n (k^2 + l^2). \tag{41}$$

The solution to Eq. (39) is

$$\delta\phi = A_0 \exp[t/\tau + il(z - V_z t)]\cos(kx), \tag{42}$$

where $A_0$ is a constant.

Eqs. (41) and (42) imply that the perturbation grows at sufficiently large $B_m$ and/or $\bar{\phi}$ and travels upward with the velocity $V_z$ that depends only on $\bar{\phi}$ (Spiegelman, 1993a). As we will discuss in detail in the next section using numerical simulation, the term proportional to $B_m$ in Eq. (41) represents the instability induced by the MMUb feedback (Ogawa, 2018): when an upwelling flow of matrix generates magma by decompression melting, its buoyancy accelerates the upwelling flow itself. On the other hand, the term that contains $3\bar{\phi}^2/\phi_0^3$ represents the excitation of the "porosity wave" (Spiegelman, 1993a, b); a perturbation in $\phi$ induces a spatial variation in the rate of volume change of matrix (Eq. (27)), leading to an upward shift of the perturbation as a wave. Both modes of instability are induced by decompression melting of the matrix: we will describe these modes in more detail in the next section using numerical models.

Following the standard procedure (e.g., Turcotte & Schubert, 2014), we looked for the value of $k$ that maximizes $\tau^{-1}$ in Eq. (41) and present the thus calculated $\tau^{-1}$ (the contour lines) together with the corresponding value of $k$ for the fastest growing mode $k_{max}$ (the color) as a function of $B_m$ and $\bar{\phi}$ in Figure 2. Here, $l$ is fixed at $6\pi$, $T_0$ at 2.57, $S_d$ at 0.1, $G$ at 0.05, and $\kappa_n$ at 0.01. (We chose this value of $l$ for the sake of a comparison of Figure 2 with the numerical simulations discussed in the next section.) We also present, by the dashed line, the values of $B_m$ and $\bar{\phi}$ at which the first and the second terms in the parenthesis of Eq. (41) are equal in Figure 2:

$$B_m = \frac{3\bar{\phi}^2}{\phi_0^3}(k_{max}^2 + l^2). \tag{43}$$

There are three regimes in the figure: (a) the regime at very small $\bar{\phi}$ and $B_m$ where the growth rate $\tau^{-1}$ is negative and the steady state expressed by Eqs. (19) to (21) is stable due to the diffusion term in Eq. (41); (b) the regime at $B_m$ above the threshold calculated from Eq. (43) where the MMUb feedback dominates the flow of magma and matrix because of the strong effect of melt buoyancy; and (c) the regime at large $\bar{\phi}$ and $B_m$ below the threshold where the porosity waves dominate the flow because the MMUb feedback is negligible. The threshold in $B_m$ is $10^5 \sim 10^6$ in Figure 2. Note that the wavelength of the fastest growing mode for the MMUb feedback $(= 2\pi/k_{max})$ is longer than that of the porosity wave. The critical values $B_m^c$ and $\bar{\phi}^c$ indicated in Figure 2 can be analytically calculated as

$$B_m^c = \frac{27}{4}\frac{(1+G)\kappa_n l^4}{T_0 S_d}, \quad \bar{\phi}^c = \sqrt{\frac{4\phi_0^3(1+G)\kappa_n l^2}{3T_0 S_d}}, \tag{44}$$

respectively. At the parameter values assumed in Figure 2, $B_m^c = 3.2 \times 10^4$ and $\bar{\phi}^c = 0.05$ hold.

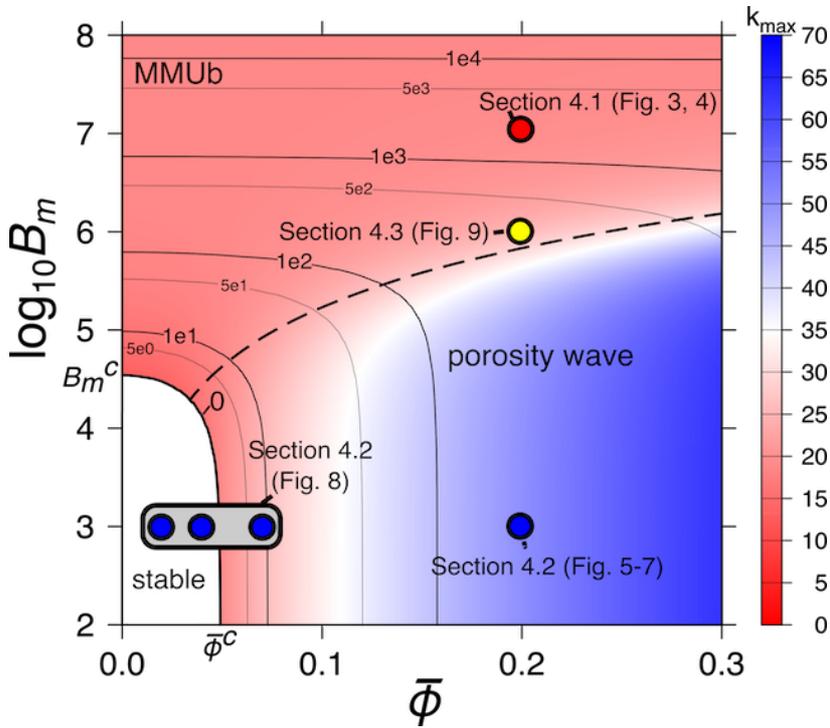

**Figure 2.** Map of the growth rate $\tau^{-1}$ (the contour lines) and the wave number of the fastest growing mode $k_{max}$ (color), plotted on the parameter space of the background melt-content $\bar{\phi}$ and melt buoyancy parameter $B_m$. The numbers at the contour lines indicate the values of $\tau^{-1}$. Other parameters are fixed as $l = 6\pi$, $T_0^* = 2.57$, $S_d = 0.1$, $G = 0.05$, and $\kappa_n = 0.01$. The thick solid line marks the boundary of the stable and unstable regimes where $\tau^{-1} = 0$. The dashed line represents the condition $B_m/(k_{max}^2 + l^2) = 3\bar{\phi}^2/\phi_0^3$: above this line the instability is dominated by the MMUb feedback, whereas below it the porosity-wave-driven instability occurs. Also plotted are the parameter values adopted in the numerical simulations discussed in Section 4.1 to 4.3. In many of the simulations, the background $\bar{\phi}$ is initially layered, not uniform as assumed in the linear perturbation analysis. In these cases, the initial value for the lower layer $\phi_{init}^l$ is plotted in the figure. For $B_m^c$ and $\bar{\phi}^c$, see the text.

## 4 Numerical simulations in a 2-D Box

To investigate the nonlinear growth of the disturbances discussed in Section 3, we carried out a series of numerical simulations of magma-migration and matrix flow in a 2-D box. The employed basic equations are presented in Section 2. The top and bottom boundaries of the box as well as the side walls are impermeable and shear stress-free. The temperature in the box is equal to the melting temperature (Eq. (15)), as in the linear analysis in Section 3, and the initial distribution of melt-content is uniform with $\phi_{init} = 0.2$ or layered with the content in the upper layer ($z > 0.5$) $\phi_{init}^u = 0.005 \sim 0.16$ and that in the lower layer ($z < 0.5$) $\phi_{init}^l = 0.02 \sim 0.2$. To start the calculation, we imposed a sinusoidal perturbation as in Section 3 to the initial melt-content with vertical wave number $l = 6\pi$ and the horizontal wave number equal to that of the most unstable mode $k = k_{max}$ shown in Figure 2; the amplitude of the perturbation is $0.1\bar{\phi}$. (We confirmed that the results are not sensitive to the choice of the amplitude.) The adopted values of other model parameters are $T_0^* = 2.57 (\equiv 1360\ K)$, $S_d = 0.1$, and $\kappa_n = 0.01$, unless otherwise mentioned; the adopted values for each case are listed in Table 2. In the following, we discuss the deviation of matrix velocity $U_d$ from the initial reference velocity calculated from $\phi_{init}$ and Eq. (20) ($U_d = U - \bar{U}_{init}$) rather than $U$ itself.

Table 2. The parameter values assumed in each case.

| Case # | $B_m$ | $\phi_{init}^u$ | $\phi_{init}^l$ | $\kappa_n$ | Figure |
|---|---|---|---|---|---|
| E7M02 | $1 \times 10^7$ | 0.2 | 0.2 | 0.01 | 3 |
| E7M0005-02 | $1 \times 10^7$ | 0.005 | 0.2 | 0.01 | 4 |
| E3M02 | $1 \times 10^3$ | 0.2 | 0.2 | 0.01 | 5 |
| E3M0005-02 | $1 \times 10^3$ | 0.005 | 0.2 | 0.01 | 6 |
| E3M016-02 | $1 \times 10^3$ | 0.16 | 0.2 | 0.01 | 7 |
| E3M01-02 | $1 \times 10^3$ | 0.1 | 0.2 | 0.01 | 7 |
| E3M0005-002 | $1 \times 10^3$ | 0.005 | 0.02 | 0.01 | 8 |
| E3M0005-004 | $1 \times 10^3$ | 0.005 | 0.04 | 0.01 | 8 |
| E3M0005-007 | $1 \times 10^3$ | 0.005 | 0.07 | 0.01 | 8 |

| Case # | $B_m$ | $\phi_{init}^u$ | $\phi_{init}^l$ | $\kappa_n$ | Figure |
|---|---|---|---|---|---|
| 007-kap | $1 \times 10^3$ | 0.005 | 0.07 | 0.01 at $\phi > 0.001$ | 8 |
| E6M0005-02 | $1 \times 10^6$ | 0.005 | 0.2 | 0.01 | 9 |

### 4.1 From the MMUb Feedback to Partially Molten Plumes

We first examine the flow of magma and matrix above the threshold for the MMUb feedback shown in Figure 2. The adopted parameter value is $B_m = 1 \times 10^7$. At this value of $B_m$, the first term on the right-hand side of Eq. (38) is larger than the second term, and the equation predicts that $\delta U_z$ should be positively correlated with $\delta \phi$: when magma is generated by upwelling flow of matrix ($\delta \phi > 0$), the buoyancy of the magma boosts the upwelling flow itself ($\delta U_z > 0$).

#### 4.1.1 The Case with Uniform Initial Melt-Distribution

Figure 3 shows the case with the uniform initial distribution of $\phi_{init} = 0.2$ (Case E7M02). At first, the matrix moves upward in the initially high-$\phi$ regions with positive perturbations of $\phi$ and downward in the initially low-$\phi$ regions with negative perturbations, as predicted by Eq. (38) (Figure 3a-2). As the perturbation evolves, four initially high−$\phi$ patches coalesce to cause a strong upwelling flow at the center of the box and descending counter flows along the sidewalls, forming a convective circulation throughout the box (Figure 3a-4): the central upwelling induces decompression melting, which generates additional magma and further strengthens the upward flow of matrix. The plot of root-mean-square (rms) velocity of the matrix indicates that this nonlinear growth becomes substantial at around a nondimensional time of $3.2 \times 10^{-4}$ (Figure 3b), which is comparable to the growth timescale predicted by the linear analysis (~$4 \times 10^{-4}$, Figure 2). Note that the descending counter flow suppresses melting in the adjacent regions in Figure 3a-4. The flow shown in Figure 3 confirms the above prediction that the MMUb feedback dominates the convective flow.

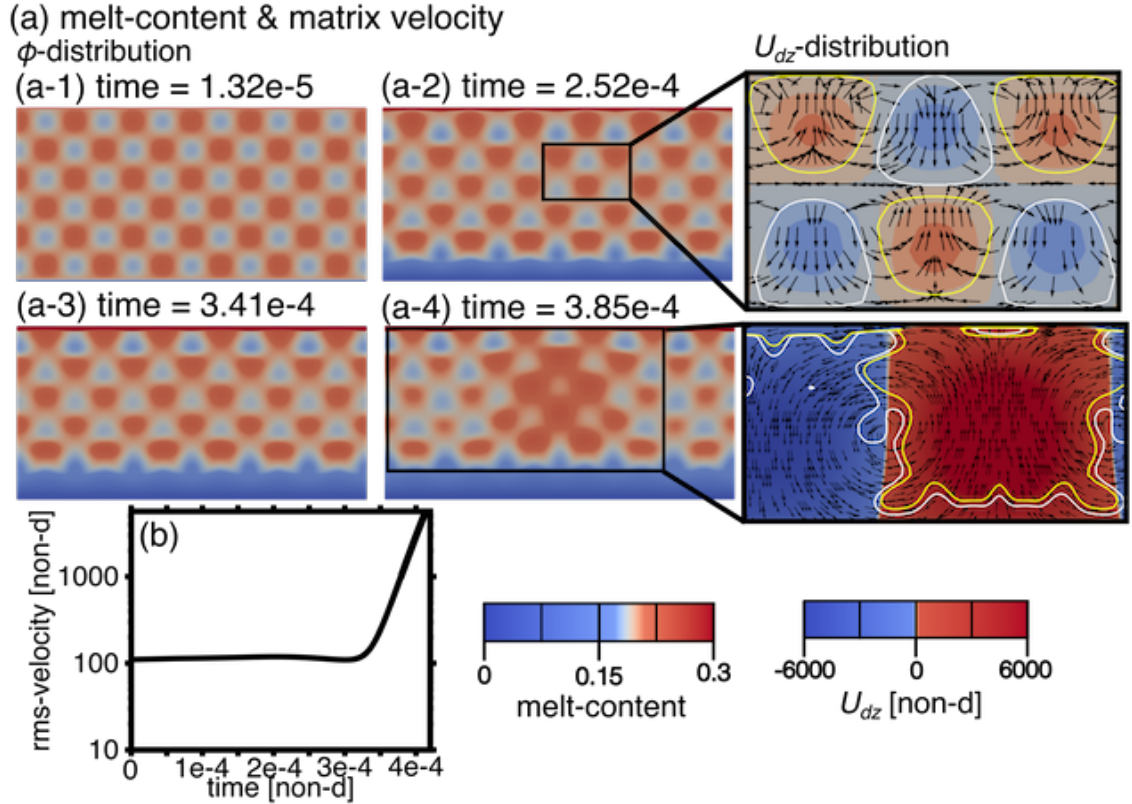

**Figure 3.** (a) Presented are snapshots of the melt-content $\phi$ and matrix velocity $\boldsymbol{U}_d$ fields for Case E7M02 where the initial melt-content distribution $\phi_{init}$ is uniform at 0.2, $B_m = 1 \times 10^7$, and $l = 6\pi$ (see Figure 2). The color and the arrows in the snapshots for $\boldsymbol{U}_d$ show its z-component $U_{dz}$ (red for upward and blue for downward) and direction, respectively. (The arrows express the direction of $\boldsymbol{U}_d$ but does not its magnitude.) The white and yellow lines represent the contours of $\phi = 0.196$ and $0.204$, respectively. Also shown is (b) the root-mean-square (rms) velocity of the matrix $\boldsymbol{U}$ plotted against time for this case.

### 4.1.2 The Case with Layered Initial Melt Distribution

When the initial melt-distribution is layered, the MMUb feedback induces partially molten plumes along the layer boundary as shown in Figure 4 and Movie 1 in the supporting information (Case E7M0005-02). Here, the parameter values adopted are the same as those in Case E7M02 (Figure 3) except that the initial melt-content in the upper layer $\phi_{init}^u = 0.005$. An initial perturbation is imposed only in the lower layer, and the assumed horizontal wavenumber for the perturbation $k$ is equal to that of the most unstable mode (Figure 2).

As in the uniform case (Figure 3), the matrix migrates upward in high-$\phi$ regions and downward in low-$\phi$ regions (Figure 4a-1). Both the matrix and melt in high-$\phi$ regions in the lower layer ascend due to melt buoyancy and intrudes into the upper layer, forming partially molten plumes along the layer boundary (see Figure 4a-2 and Movie 1). The rms-velocity gradually increases throughout the calculation (Figure 4b) at a rate higher than that in the uniform case (Figure 3b) because the perturbation in the lower layer induces partially molten plumes from the very beginning of the calculation (Figure 4a-1).

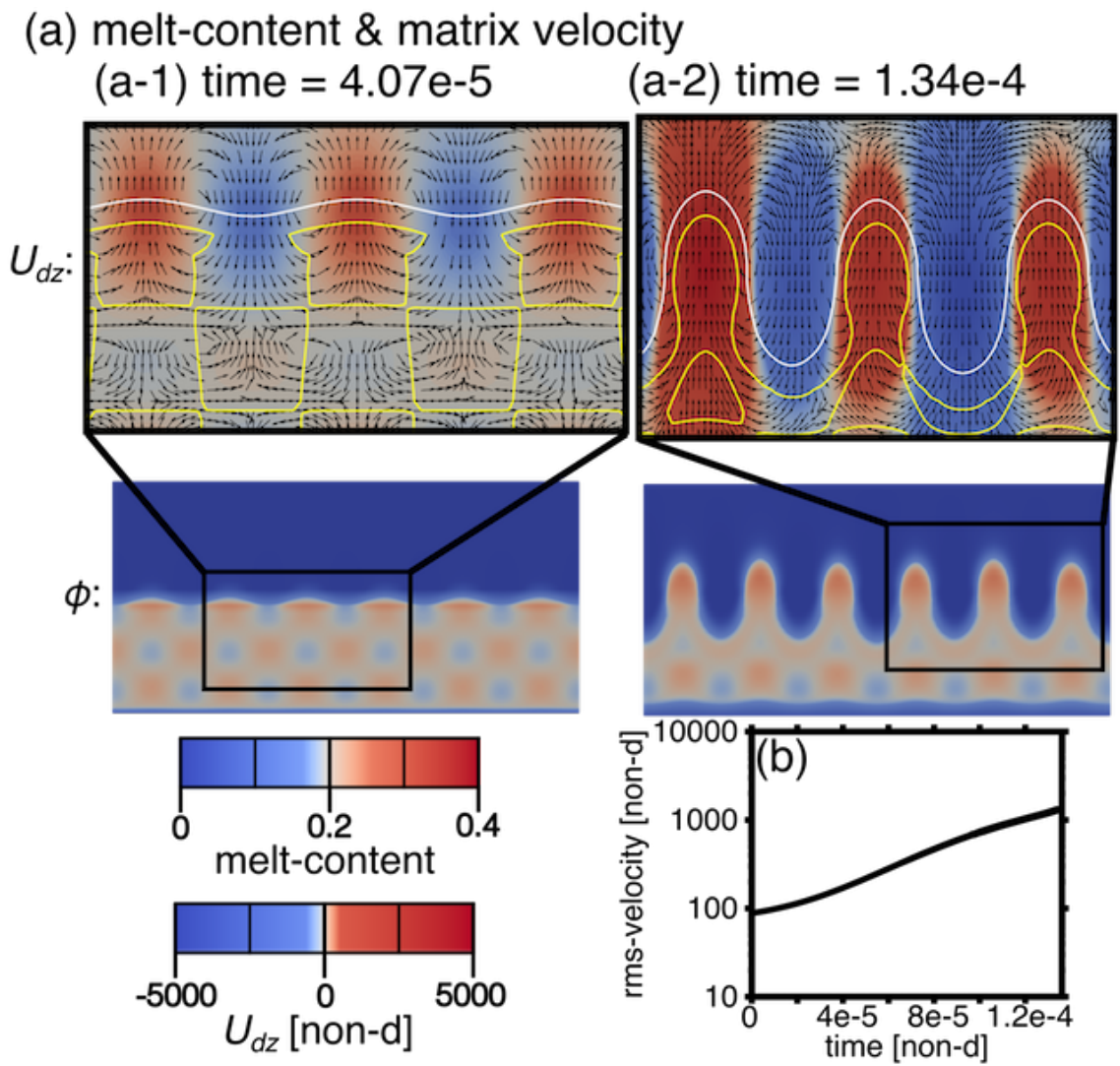

**Figure 4.** The same as Figure 3 but for Case E7M0005-02 where the initial melt-content in the upper layer ($z > d/2$) $\phi_{init}^{u}$ is 0.005 and that in the lower layer ($z < d/2$) $\phi_{init}^{l}$ is 0.2. Other parameter values are the same as those of Figure 3. The white and yellow contour lines represent the $\phi$-distribution at 0.1 and 0.2, respectively.

## 4.2 From Porosity Wave to Melt-Finger

To examine the porosity-wave regime shown in Figure 2, we next calculated several cases at $B_m = 1 \times 10^3$. At this value of $B_m$, the second term on the right-hand side of Eq. (38) is larger than the first term, and $U_{dz}$ ($=\delta U_z$) should be negatively correlated with $\delta\phi$.

### 4.2.1 The Case with Uniform Initial Melt-Distribution

Figure 5 shows the results we obtained when the initial melt-content is uniform at $\phi_{init} = 0.2$ (Case E3M02). In contrast to the case with $B_m = 1 \times 10^7$ shown in Figure 3, the matrix migrates downward in the high-$\phi$ regions with positive perturbations and upward in the low-$\phi$ regions with negative perturbations as predicted from the minus sign of the second term on the right-hand side of Eq. (38). The downward flows of matrix in high-$\phi$ regions where the upward migration of magma is enhanced are caused by the combination of the expansion and contraction of matrix at the top and bottom of the high-$\phi$ region, respectively (see the thick blue circles in Figure 5a and Eq. (27)). The upward phase velocity of the porosity wave is approximately 600, in agreement with the value of $V_z$ calculated from Eq. (40). (The high-$\phi$ layer along the top boundary and the almost melt-free layer along the bottom boundary of the box develop owing to the background upward flow of magma described by Eq. (21).)

In contrast to the MMUb feedback discussed above, nonlinear behavior of the porosity wave is not observed in Figure 5. This is because the growth rate of the porosity wave by decompression melting is around 300 (see Figure 2) and the growth is not conspicuous on the timescale required for magma to migrate from the bottom to the top boundaries ($\sim \phi_0^3/\phi^2 = 3.1 \times 10^{-3}$, see Eq. (21)).

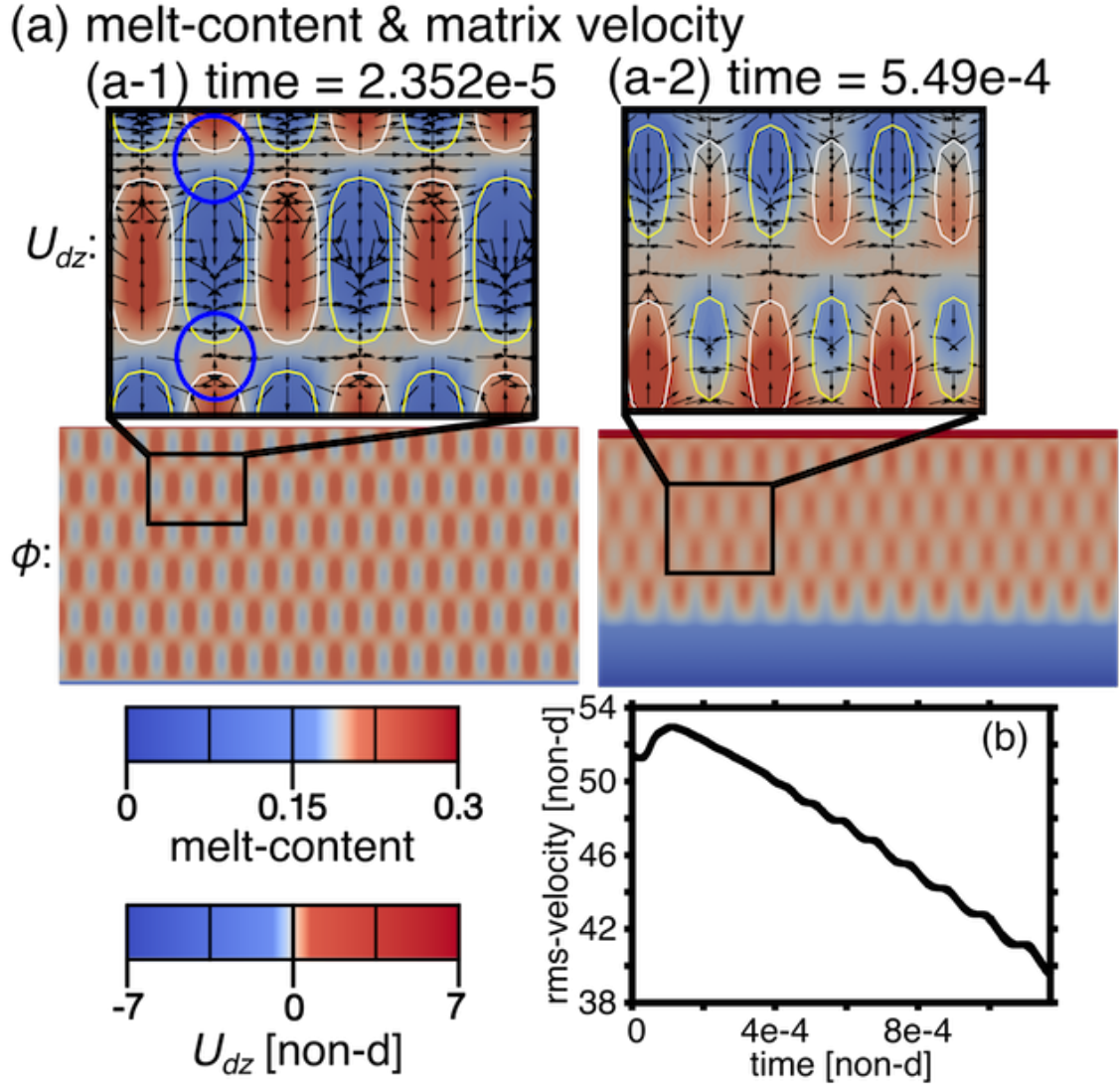

**Figure 5.** The same as Figure 3 but for Case E3M02 where the melt buoyancy number $B_m$ is $1 \times 10^3$ (see the blue dot in Figure 2). The thin white and yellow lines represent the contours of $\phi = 0.196$ and $0.204$, respectively. For the thick blue circles, see the text.

### 4.2.2 The Cases with Layered Initial Melt Distribution

When the initial melt-distribution is layered, the porosity wave in the lower layer induces melt-fingers along the layer boundary as shown in Figure 6 and Movie 2 (Case E3M0005-02): the initial melt-content in the upper layer is $\phi_{init}^u = 0.005$ and that in the lower layer is $\phi_{init}^l = 0.2$. The upward migrating melt in the lower layer accumulates along the layer boundary because of the lower permeability in the upper layer due to the lower $\phi_{init}^u$, forming a continuous layer (Figure 6a-1 to 6a-2). This melt-enriched layer undulates as the porosity wave in the lower layer propagates upward to collide with the layer boundary (Figure 6a-2). Once most of the melt in the lower layer accumulates in the melt-enriched layer, the divergent flow of matrix around the crests of the undulation causes a downward flow that depresses the partially molten materials around the adjacent troughs to suppress melting there, further amplifying the undulation (see the downward matrix flow in the yellow circle in Figure 6a-3). This instability of melt-enriched layer causes a strong flow of matrix after $t \approx 8 \times 10^{-4}$ (Figure 6b). Eventually, the layer is disintegrated into isolated patches of melt that rise upward (Figure 6a-4 and Movie 2). These upwelling patches of melt are referred to as "melt-fingers" in U et al. (2023, 2025).

To more clearly understand the growth of melt-fingers, we calculated two cases where the initial melt-content in the upper layer $\phi_{init}^u = 0.16$ (Case E3M016-02) and $\phi_{init}^u = 0.1$ (Case E3M01-02); other parameters are the same as those assumed in Case E3M0005-02 shown in Figure 6. When $\phi_{init}^u = 0.16$, the contrast in melt-velocity between the upper and lower layers is too small for melt to accumulate along the layer boundary, and the perturbation in the lower layer simply propagates upward (Figure 7a). When $\phi_{init}^u = 0.1$, on the other hand, magma does accumulate along the boundary (Figure 7b). The melt-enriched layer is, however, diffuse and does not further develop into melt-fingers, resulting only in a wavy layer boundary. As a consequence, the rms-velocity of matrix in Figure 7c fluctuates only slightly for the cases with $\phi_{init}^u = 0.16$ (the light blue line) and 0.1 (the blue line) owing to the porosity wave. The much larger fluctuation in the rms-velocity observed in Figure 6b suggests that melt-fingers are a consequence of an instability of the thin melt-enriched layer caused by the porosity wave and are not a simple manifestation of the porosity wave truncated at the layer boundary.

Additional calculation of several cases with different initial melt-content in the lower layer $\phi_{init}^l$ also helps us understand the nature of melt-fingers. As can be seen from Figure 8, melt-fingers do not grow when $\phi_{init}^l$ is in the stable regime shown in Figure 2; there is no sign of melt-fingers at $\phi_{init}^l = 0.02$ and $0.04$ (Figure 8a, b). In contrast, a slight undulation of the layer boundary is observed at $\phi_{init}^l = 0.07$ (Figure 8a, c). Melt-fingers, however, are not observed in this case despite that the lower layer is in the regime of porosity wave (Figure 2). To understand why melt-fingers grow at $\phi_{init}^l = 0.2$ but do not at $\phi_{init}^l = 0.07$, we calculated Case 007-kap where the diffusion in the energy equation Eq. (13) is considered only in the lower layer: $\kappa_n$ is assumed to be 0.01 when $\phi > 0.001$ but 0 when $\phi < 0.001$ (Figure 8d). Compared with the case shown in Figure 8c (E3M0005-007), a thin melt-enriched layer develops along the top of the lower layer due to the weaker effect of diffusion and undulate to develop melt-fingers; the rms-velocity fluctuates with a larger amplitude, accordingly (Figure 8a). This result reinforces the above suggestion that a melt-enriched layer along the layer boundary is the source of melt-fingers.

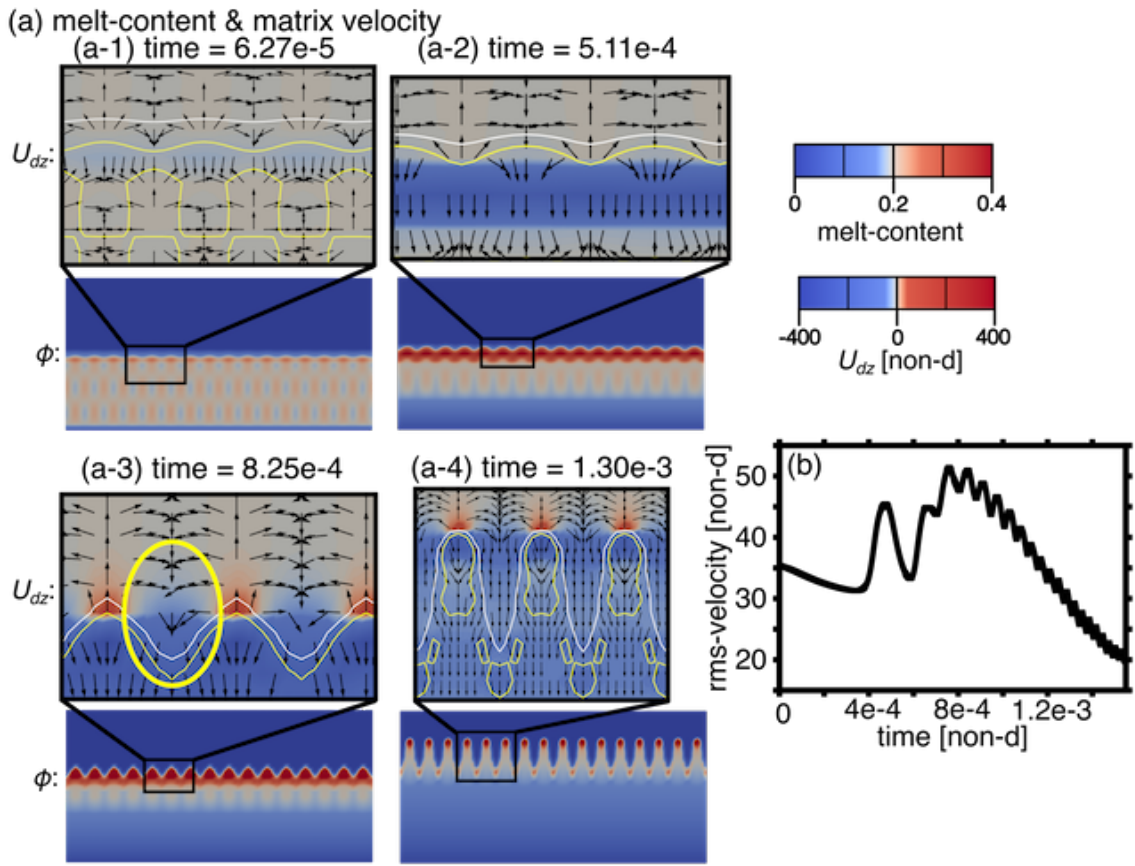

**Figure 6.** The same as Figure 4 but for the layered Case E3M0005-02 ($B_m = 1 \times 10^3$, $k = 16\pi$) where $\phi_{init} = 0.005$ at $z > d/2$ and $\phi_{init} = 0.2$ at $z < d/2$. The thin white and yellow lines represent the contours of $\phi = 0.1$ and $0.2$, respectively. For the thick yellow circles, see the text.

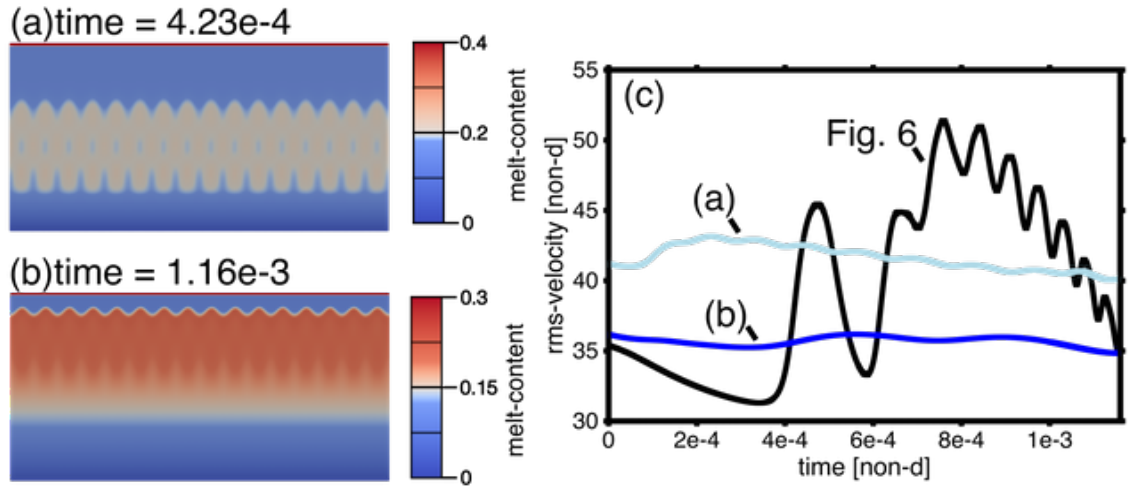

**Figure 7.** The same as Figure 6a but for (a) Case E3M016-02 and (b) Case E3M01-02 where the initial melt-content $\phi_{init}$ is 0.16 and 0.1 in the upper layer ($z > d/2$), respectively. (c) rms-velocity of the matrix plotted against time for Case E3M016-02 (the light blue line), Case E3M01-02 (the blue line), and Case E3M0005-02 (the black line taken from Figure 6). The initial melt content in the lower layer $\phi_{init}^l$ is 0.2 in all of the cases.

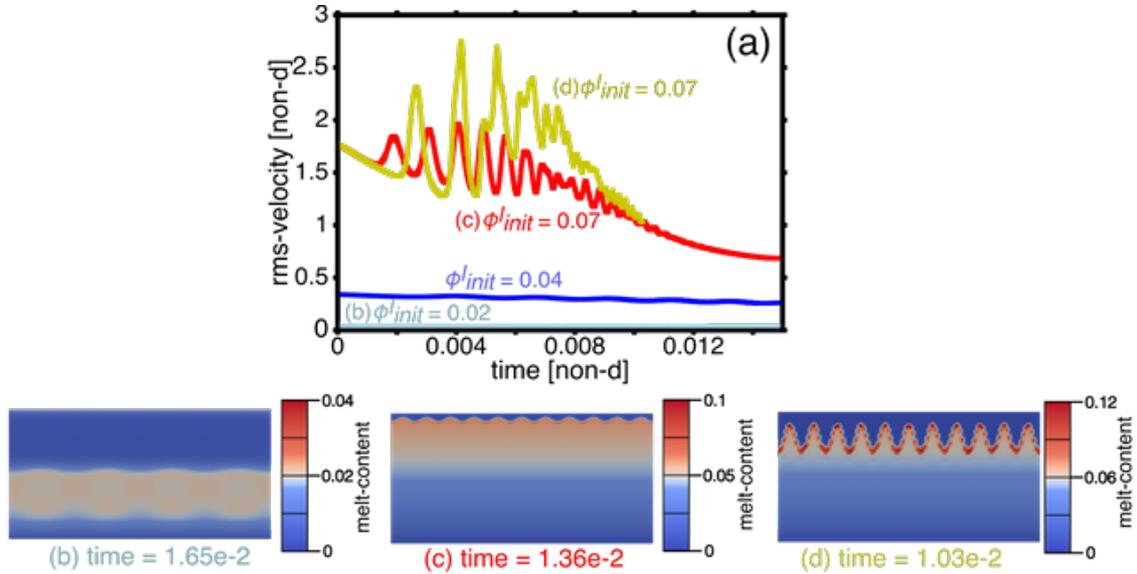

**Figure 8.** (a) Rms-velocity of the matrix plotted against time for E3M0005-002 (the light blue line), Case E3M0005-004 (the blue line), and Case E3M0005-007 (the red line) where the initial melt-content in the lower layer ($z < d/2$) $\phi_{init}^l$ is assumed to be 0.02, 0.04, and 0.07, respectively. Also shown is Case 007-kap (the yellow line); the same as Case E3M0005-007 but for the case where the diffusion is considered only in the lower layer. The snapshot of $\phi$-distributions are for (b) Case E3M0005-002, (c) Case E3M0005-007, and (d) Case 007-kap indicated in (a).

## 4.3 From Melt-Fingers to Partially Molten Plumes

At around the regime boundary between the MMUb feedback and the porosity wave (Figure 2), melt-fingers first develop and then evolve into partially molten plumes in the calculation with the layered initial condition, as shown in Figure 9 and Movie 3. (Here, we assumed that $B_m = 1 \times 10^6$ and other parameter values the same as those in Case E3M0005-02 shown in Figure 6.) In this case, perturbations in $\phi$-distribution grow along the layer boundary as a melt-enriched layer develops there (Figure 9a-1). The perturbations develop into melt-fingers at $t = 4.43 \times 10^{-4}$ (Figure 9a-2); the matrix in the crests of the perturbations moves downward. The rms-velocity is close to that of Case E3M0005-02 shown in Figure 6b at this stage (see the black and red lines in Figure 9b). As the melt-fingers grow, however, the matrix velocity field gradually shifts from downward to upward within the melt-fingers, which turn into partially molten plumes (see Figure 9a-3 and Movie 3). Finally, these plumes cause a box-wide convective flow that is much stronger than the flow induced by melt-fingers (Figure 9a-4 and 9b).

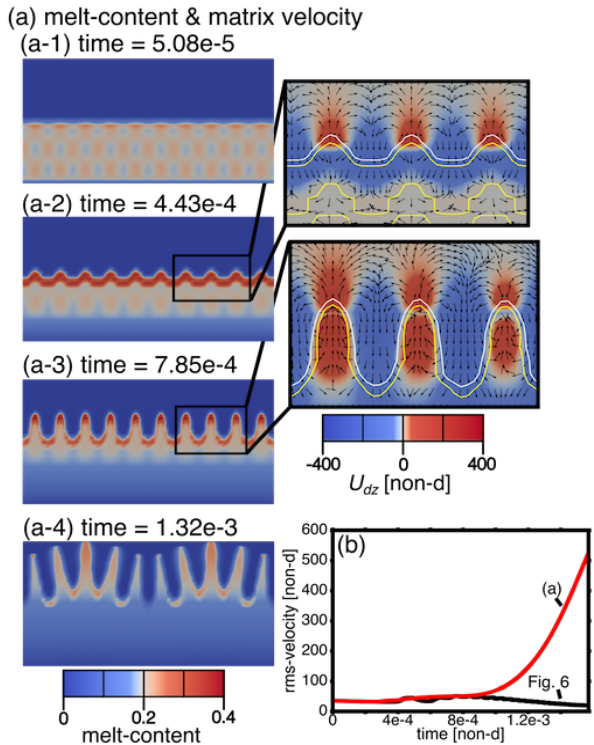

**Figure 9.** The same as Figure 6 but for Case E6M0005-02 where $B_m$ is $1 \times 10^6$. In (b), we also plot the rms-velocity of Case E3M0005-02 taken from Figure 6. The white and yellow contour lines represent the $\phi$-distribution at 0.1 and 0.2, respectively.

# 5 Discussion

We carried out a linear perturbation analysis and a series of numerical experiments to better understand the dynamics of partially molten mantle at the length scale of hundreds of kilometers and found that magma in a partially molten region in the deep mantle can migrate upward through the overlying solid mantle in two modes. When the melt-buoyancy parameter $B_m$ defined by Eq. (12) is higher than a threshold calculated from Eq. (43), the magmatism–mantle upwelling feedback (MMUb) dominates the upward magma migration (Figure 2). Partially molten materials ascend from the partially molten region through the solid mantle as plumes driven by melt-buoyancy while more melt is generated in the plumes by the ascending flow (Figure 4). At lower $B_m$, in contrast, the porosity wave in the partially molten region causes an undulation of a melt-enriched layer that develops along the top of the region and fragments the layer into isolated patches of melt. The patches ascend through the overlying solid mantle as melt-fingers (Figure 6).

## 5.1 Comparison with Previous Studies

The MMUb feedback has already been noticed, as mentioned in Section 1 (e.g., Tackley & Stevenson, 1993; Jha et al., 1994; Ogawa, 2018, 2020). The threshold for the MMUb feedback to be dominant, Eq. (43), is consistent with that empirically obtained by Ogawa (2018). (There is an error in Ogawa (2018); Eq. (43) provides the correct threshold.) This threshold is different from the one derived in Hernlund et al. (2008b) who discussed migration of magma generated by upwelling mantle flow beneath ocean ridges. Their criterion (Figure 5 in Hernlund et al., 2008b) shows the condition that the mantle flow driven by melt buoyancy becomes significant compared with the upwelling flow induced by diverging tectonic plates. Eq. (43), in contrast, shows the threshold above which the flow of matrix driven by the MMUb feedback becomes more important than that driven by the porosity wave.

The melt-fingers that we calculated here are different from most of the various types of localized melt and melt channels that develop on the spatial scale of a kilometer or less by compaction of matrix, as already mentioned in Section 1 (e.g., Aharonov et al., 1995; Connolly and Podladchikov, 2007; Bessat et al., 2022). The solitary wave of melt found in Richter & Daly (1989) is, however, close to the melt-fingers discussed here. In their analyses, Richter & Daly (1989) retained the contribution of matrix-compaction in the percolation equation (see their Eq. (2)). On the spatial scale of several tens of kilometers employed in their numerical experiments, however, this contribution is negligible (Ribe, 1985) and their percolation equation is practically identical to our Eq. (4). Although Richter & Daly (1989) calculated migration of magma generated by a prescribed magma-source, here we self-consistently calculated magma-generation by the decompression melting that the matrix flow caused by the migrating magma itself induces (Eq. (5)) and still found growth of melt-fingers. Obviously, the melt-fingers are different from the viscous fingers that arise when the viscosity of matrix is highly variable (e.g., Homsy, 1987; Kim, 2022).

## 5.2 Implications

Our analyses indicate that the melt-fingers observed in Figure 1 and U et al. (2023, 2025) are caused by the porosity wave. Indeed, we calculated the model of Figure 1 at $d = 1350$ km, $\eta = 10^{20}$ Pa s, $k_{\phi_0} = 8.6 \times 10^{-15}$ m$^2$, and $\mu = 1$ Pa s, which leads to $B_m = 2.1 \times 10^6$ for the partially molten layer at the base of the mantle where $\bar{\phi}$ is 0.1 to 0.25 shown in Figure 1. In Figure 2, these parameter values are close to the boundary between the regime of the MMUb feedback and that of the porosity wave (see the dashed line in Figure 2). The closeness is the reason why melt-fingers develop first to trigger partially molten plumes later in U et al. (2023, 2025), as is the case for Figure 9.

Melt-fingers are a salient feature of the model of mantle evolution in our model for the Moon but have not been observed in the models of larger planets like Mars, Venus, and the Earth despite that the same basic equations are employed (Ogawa, 2014; Ogawa and Yanagisawa, 2014; Ogawa, 2021). The difference arises from the pressure-dependence of the density contrast between magma and matrix assumed in these models. In this study, we assume that magma is always less dense than the matrix (Eq. (3)). For a peridotite mantle, this assumption does not necessarily hold; a density inversion is suggested to occur between the solid mantle and magma at pressure greater than around 10 GPa (e.g., Agee, 1998; Suzuki and Ohtani, 2003; Sakamaki et al., 2010). This pressure is higher than that at the base of the mantle in Mercury and the Moon (Garcia et al., 2011, 2012; Xu et al., 2024), and melt-fingers can develop even at the base of the mantle in these terrestrial bodies (see Figure 5 in Ogawa, 2016; see also Figure 2 in U et al., 2023). In larger planets, however, the assumption holds only in the shallow mantle: at depths of less than around 800 km in Mars and around 300 km in Venus and the Earth (e.g., Aitta, 2012; Yoshizaki and McDonough, 2020). The mantle is strongly stirred there by the convective flow driven by the MMUb feedback in the models of e.g., Ogawa (2021), and the growth of melt-fingers is inhibited. Melt-fingers can readily develop only in small terrestrial bodies like the Moon.

The threshold for the MMUb feedback expressed by Eq. (43) has an implication for the Earth's volcanic activity. This equation implies that the MMUb feedback is important when the size of partially molten region $L$ is larger than the threshold calculated from

$$L > \sqrt{24\pi^2 \frac{\bar{\phi}^2}{\phi_0^3} \frac{\eta k_{\phi_0}}{\mu}}. \qquad (45)$$

Here, we assume $k_{max} \sim l$ for simplicity. When $\bar{\phi} = 0.1$, $\eta = 10^{20}$ Pa s, $k_{\phi_0} = 10^{-14}$ m$^2$ (Miller et al., 2014), and $\mu = 1$ Pa s (see Table 1), Eq. (45) implies that the MMUb feedback operates when $L > 138$ km. In the Earth, plume heads beneath hotspots are typically ~100 km in radius (e.g., Wolfe et al., 1997; Steinberger and Antretter, 2006), marginally less than the threshold, and the MMUb feedback is probably not important for the hotspot volcanic activity. The sizes of the partially molten sources for large igneous provinces (LIPs) are, in contrast, estimated to be around 300 km for the Columbia River basalts to around 1000 km for the Ontong-Java plateau (Coffin and Eldholm, 1994; Sobolev et al., 2011), both above the threshold. The difference in spatial scale is, thus, the likely reason for the difference in the style of volcanic activities between hot spots and LIPs, and the massive eruptions of plateau basalts at LIPs are most likely to be a consequence of the MMUb feedback. (Note that the MMUb feedback operates only in the depth-range less than around 300 km in the Earth because of the melt-solid density inversion at greater depths, as mentioned above. What we observed in our model of the Earth's mantle dynamics (e.g., Ogawa, 2014) is that the MMUb feedback that operates in the uppermost mantle causes a convective flow that extends to a greater depth, and this flow transports the hot mantle materials there upward to the site of the feedback to boost the feedback itself.)

There are no observational constraints on the size of the source regions of volcanic activities on other planets like Mars where both hot spots and LIPs are present (e.g., Grott et al., 2013; Krishnan and Kumar, 2023). An analogy from the Earth, however, suggests that the spatial scale that determines whether or not the MMUb feedback operates is a possible key to the distinction between these two types of volcanism. The threshold (Eq. (45)) implies that LIPs were more important in the earlier history of a planet because of the expected higher temperature in its mantle; a higher temperature reduces the threshold by reducing the viscosity in the mantle and also makes partially molten region in the mantle larger. This is what is observed for Mars (Grott et al., 2013) and also in our models of mantle evolution in Mars, Venus, and the Earth (Ogawa, 2014; Ogawa & Yanagisawa, 2014; Ogawa, 2021). (On the Moon, there are only LIPs and no hot spots have been reported (e.g., Hiesinger et al., 2003; Morota et al., 2011; Head et al., 2023). This is probably because the lunar mantle is below the threshold for onset of thermal convection by basal heating (Yanagisawa et al., 2016) and melt-buoyancy is the only possible driving mechanism of mantle convection, as assumed in our model (U et al., 2023); there is likely no hot plume driven by thermal buoyancy that causes hot spot volcanism in the Moon.)

# 6 Conclusion

To understand the dynamics of a partially molten mantle on the length scale of hundreds of kilometers, we carried out a linear perturbation analysis and 2-D numerical simulations of a convective flow in a partially molten horizontal layer where magma migrates upward through the coexisting mantle as a permeable flow.

The linear perturbation analysis shows that there are two modes in the upward migration of magma (Figure 2). When the melt-buoyancy parameter $B_m$ defined by Eq. (12) exceeds a threshold, the magmatism–mantle upwelling feedback (MMUb) dominates magma migration (Figure 2 and Eq. (43)): an upward flow of matrix generates magma by decompression melting, while the buoyancy of the generated magma boosts the original upward flow. When a solid layer is overlaid on the partially molten layer, the MMUb feedback induces partially molten plumes that ascend through the solid layer owing to their melt buoyancy (Figure 4a-2). At lower $B_m$, in contrast, the porosity wave dominates the flow of magma. A perturbation in the distribution of melt-content propagates upward because it induces a perturbation in the distribution of upward velocity of magma, which in turn causes a variation in the rate of volume expansion or contraction of matrix. When the partially molten layer is overlaid by a solid layer, the porosity wave causes an instability along the top of the partially molten layer to form patches of magma, or melt-fingers, which ascend through the solid layer (Figure 6a-4). Near the regime boundary between the MMUb feedback and the porosity wave, melt-fingers initially develop and subsequently evolve into partially molten plumes (Figure 9).

At a typical value of the viscosity of matrix appropriate for the upper mantle of the Earth, $10^{20}$ Pa s, the MMUb feedback is substantial only when the spatial scale of a partially molten region is larger than around 100-300 km (see Eq. (45)). This threshold suggests that the feedback drives the volcanism that has formed the Large Igneous Provinces while that it is not important for the hot spot volcanism on today's Earth. Since the threshold decreases with decreasing matrix viscosity, LIPs by the MMUb feedback are likely to have been more important in the earlier Earth, where the mantle was hotter and softer, and by analogy in the early history of Mars and other terrestrial planets. Melt-fingers are, in contrast, expected to have developed in the Moon where magma is less dense than the coexisting matrix in the entire, or at least in most part of, the mantle, if a partially molten layer has developed at its base.

**Declarations**

**Ethics approval and consent to participate**

Not applicable

**Consent for publication**

Not applicable

**Availability of data and materials**
The data that support the findings of this study and the numerical programs used in this study are available upon reasonable request to K.U. (email: kenyo.u.grc@gmail.com ).
**Competing interests**
The author declares no competing interests.
**Funding**

This work was supported by JST SPRING Grant JP-MJSP2108 and JSPS KAKENHI Grant JP202412823 of Japan.

**Authors' contributions**

K.U. and M.O. designed the project. M.K. and K.U. developed numerical codes. K.U., T.M., and T.Y. performed numerical simulations and analyzed simulation data. All authors contributed to the discussion and interpretation of the results. K.U. prepared the initial draft of the manuscript. All authors revised the manuscript.

**Acknowledgements**

This work was supported by: the Joint Usage/Research Center PRIUS at Ehime University, the Earth Simulator of Japan Agency for Marine-Earth Science and Technology (JAMSTEC), "Exploratory Challenge on Post-K Computer" (Elucidation of the Birth of Exoplanets [Second Earth] and the Environmental Variations of Planets in the Solar System), and "Program for Promoting Research on the Supercomputer Fugaku" (Toward a unified view of the universe: from large scale structures to planets). This work used computational resources of the supercomputer Fugaku provided by the RIKEN Center for Computational Science through the HPCI System Research Project (Project ID: hp230204, hp240219, and hp250226). Animations and some figures were drawn with the ParaView by Sandia National Laboratory, Kitware Inc., and Los Alamos National Laboratory.

**References**

Agee, C.B. (1998) Crystal-liquid density inversions in terrestrial and lunar magmas. Phys Earth Planet Inter 107:63–74. doi:10.1016/S0031-9201(97)00124-6

Aharonov, E., Whitehead, J.A., Kelemen, P.B., Spiegelman, M. (1995) Channeling instability of upwelling melt in the mantle. J Geophys Res Solid Earth 100(B10):20433–20450. doi:10.1029/95JB01307

Aharonov, E., Spiegelman, M., Kelemen, P. (1997) Three-dimensional flow and reaction in porous media: Implications for the Earth's mantle and sedimentary basins. J Geophys Res Solid Earth 102(B7):14821–14833. doi:10.1029/97JB00996

Aitta, A. (2012) Venus' internal structure, temperature and core composition. Icarus 218(2):967–974.

doi:10.1016/j.icarus.2012.01.007

Bessat, A., Pilet, S., Podladchikov, Y.Y., Schmalholz, S.M. (2022) Melt migration and chemical differentiation by reactive porosity waves. Geochem Geophys Geosyst 23(2):e2021GC009963. doi:10.1029/2021GC009963

Boukaré, C.E., Ricard, Y. (2017) Modeling phase separation and phase change for magma ocean solidification dynamics. Geochem Geophys Geosyst 18(9):3385–3404. doi:10.1002/2017GC006902

Breuer, D., Moore, W.B. (2015) 10.08 - Dynamics and Thermal History of the Terrestrial Planets, the Moon, and Io. In Schubert G. (ed) Treatise on Geophysics (Second Edition), pp 255–305. doi: 10.1016/B978-0-444-53802-4.00173-1

Coffin, M.F., Eldholm, O. (1994) Large igneous provinces: crustal structure, dimensions, and external consequences. Rev Geophys 32(1):1–36. doi:10.1029/93RG02508

Connolly, J.A.D., Podladchikov, Y.Y. (2007) Decompaction weakening and channeling instability in ductile porous media: Implications for asthenospheric melt segregation. J Geophys Res Solid Earth 112(B10). doi:10.1029/2005JB004213

Davies, G.F. (2011) Mantle convection for geologists. Cambridge Univ Press. doi:10.1017/CBO9780511973413

Garcia, R.F., Gagnepain-Beyneix, J., Chevrot, S., Lognonné, P. (2011) Very preliminary reference Moon model. Phys Earth Planet Inter 188(1–2):96–113. doi:10.1016/j.pepi.2011.06.015

Garcia, R.F., Gagnepain-Beyneix, J., Chevrot, S., Lognonné, P. (2012) Erratum to "Very Preliminary Reference Moon Model", by RF Garcia, J. Gagnepain-Beyneix, S. Chevrot, P. Lognonné [Phys Earth Planet Inter 188 (2011) 96–113]. Phys Earth Planet Inter 202:89–91. doi:10.1016/j.pepi.2011.06.015

Grott, M., Baratoux, D., Hauber, E., Sautter, V., Mustard, J., Gasnault, O., Toplis, M.J. (2013) Long-term evolution of the Martian crust-mantle system. Space Sci Rev 174(1):49–111. doi:10.1007/s11214-012-9948-3

Head, J.W., Wilson, L., Hiesinger, H., van der Bogert, C., Chen, Y., Dickson, J.L., Ziyuan, O. (2023) Lunar mare basaltic volcanism: Volcanic features and emplacement processes. Rev Mineral Geochem 89(1):453–507. doi:10.2138/rmg.2023.89.11

Hernlund, J.W., Tackley, P.J., Stevenson, D.J. (2008a) Buoyant melting instabilities beneath extending lithosphere: 1. Numerical models. J Geophys Res Solid Earth 113(B4). doi:10.1029/2006JB004862

Hernlund, J.W., Stevenson, D.J., Tackley, P.J. (2008b) Buoyant melting instabilities beneath extending lithosphere: 2. Linear analysis. J Geophys Res Solid Earth 113(B4). doi:10.1029/2006JB004863

Hiesinger, H., Head J.W., Wolf U., Jaumann R., Neukum G. (2003) Ages and stratigraphy of mare basalts in Oceanus Procellarum, Mare Nubium, Mare Cognitum, and Mare Insularum. J Geophys Res Planets 108(E7). doi:10.1029/2002JE001985

Homsy, G.M. (1987) Viscous fingering in porous media. Annu Rev Fluid Mech 19(1):271–311

Jha, K., Parmentier, E.M., Morgan, J.P. (1994) The role of mantle-depletion and melt-retention buoyancy in spreading-center segmentation. Earth Planet Sci Lett 125(1–4):221–234. doi:10.1016/0012-821X(94)90217-8

Kameyama, M. (2022). Numerical experiments on thermal convection of highly compressible fluids with variable viscosity and thermal conductivity in 2-D cylindrical geometry: implications for mantle convection of super-Earths. Geophys J Int 231(2), 1457–1469. doi: 10.1093/gji/ggac259

Kameyama, M. (2025) 2-D numerical experiments of thermal convection of highly viscous fluids under strong adiabatic compression: implications on mantle convection of super-Earths with various sizes. Earth Planets Space 77(1):13. doi:10.1186/s40623-025-02134-8

Katz, R.F. (2008) Magma dynamics with the enthalpy method: Benchmark solutions and magmatic focusing at mid-ocean ridges. J Petrol 49(12):2099–2121. doi:10.1093/petrology/egn058

Katz, R.F., Spiegelman, M., Langmuir, C.H. (2003) A new parameterization of hydrous mantle melting.

Geochem Geophys Geosyst 4(9). doi:10.1029/2002GC000433

Keller, T., May, D.A., Kaus, B.J. (2013) Numerical modelling of magma dynamics coupled to tectonic deformation of lithosphere and crust. Geophys J Int 195(3):1406–1442. doi:10.1093/gji/ggt306

Keller, T., Katz, R.F. (2016) The role of volatiles in reactive melt transport in the asthenosphere. J Petrol 57(6):1073–1108. doi:10.1093/petrology/egw030

Kim, M.C. (2022) Effect of nonlinear drag on the onset and the growth of the miscible viscous fingering in a porous medium. Korean J Chem Eng 39(3):548–561. doi:10.1007/s11814-021-0954-6

Kohlstedt, D.L., Hansen, L. (2015). 2.18 constitutive equations, rheological behavior, and viscosity of rocks. In Kohlstedt, D.L. (ed) Treatise on Geophysics (Second Edition), pp 441–472.

Krishnan, V., Kumar, P.S. (2023) Long-lived and continual volcanic eruptions, tectonic activity, pit chains formation, and boulder avalanches in Northern Tharsis Region: Implications for late Amazonian geodynamics and seismo-tectonic processes on Mars. J Geophys Res Planets 128(1):e2022JE007511. doi:10.1029/2022JE007511

Lourenço, D.L., Rozel, A.B., Gerya, T., et al. (2018) Efficient cooling of rocky planets by intrusive magmatism. Nat Geosci 11:322–327. doi:10.1038/s41561-018-0094-8

Lourenço, D.J., Rozel, A.B., Ballmer, M.D., Tackley, P.J. (2020) Plutonic-squishy lid: A new global tectonic regime generated by intrusive magmatism on Earth-like planets. Geochem Geophys Geosyst 21:e2019GC008756. doi:10.1029/2019GC008756

McKenzie, D. (1984) The generation and compaction of partially molten rock. J Petrol 25(3):713–765. doi:10.1093/petrology/25.3.713

Mei, S., Bai, W., Hiraga, T., Kohlstedt, D.L. (2002). Influence of melt on the creep behavior of olivine–basalt aggregates under hydrous conditions. Earth Planet Sci Lett 201(3-4), 491–507. doi: 10.1016/S0012-821X(02)00745-8

Miller, K.J., Zhu, W.L., Montési, L.G., Gaetani, G.A. (2014) Experimental quantification of permeability of partially molten mantle rock. Earth Planet Sci Lett 388:273–282. doi:10.1016/j.epsl.2013.12.003

Miyagoshi, T., Kameyama, M., Ogawa, M. (2018). Effects of adiabatic compression on thermal convection in super-Earths of various sizes. Earth Planets Space, 70(1), 200. doi: 10.1186/s40623-018-0975-5

Moore, W.B., Webb, A.A.G. (2013) Heat-pipe Earth. Nature 501:501–505. doi:10.1038/nature12473

Morota, T., Haruyama, J., Ohtake, M., Matsunaga, T., Kawamura, T., Yokota, Y., LISM Working Group (2011) Timing and duration of mare volcanism in the central region of the northern farside of the Moon. Earth Planets Space 63(1):5–13. doi:10.5047/eps.2010.02.009

Ogawa, M. (1994) Effects of chemical fractionation of heat-producing elements on mantle evolution inferred from a numerical model of coupled magmatism-mantle convection system. Phys Earth Planet Inter 83(2):101–127. doi:10.1016/0031-9201(94)90067-1

Ogawa, M. (2003) Chemical stratification in a two-dimensional convecting mantle with magmatism and moving plates. J Geophys Res Solid Earth 108(B12). doi:10.1029/2002JB002205

Ogawa, M. (2014) A positive feedback between magmatism and mantle upwelling in terrestrial planets: Implications for the Moon. J Geophys Res Planets 119(11):2317–2330. doi:10.1002/2014JE004717

Ogawa, M. (2018) Magmatic differentiation and convective stirring of the mantle in early planets: the effects of the magmatism-mantle upwelling feedback. Geophys J Int 215(3):2144–2155. doi:10.1093/gji/ggy413

Ogawa, M. (2020) Magmatic differentiation and convective stirring of the mantle in early planets–2: effects of the properties of mantle materials. Geophys J Int 220(2):1409–1420. doi:10.1093/gji/ggz499

Ogawa, M. (2021) The four-stage evolution of Martian mantle inferred from numerical simulation of the magmatism-mantle upwelling feedback. J Geophys Res Planets 126(12):e2021JE006997. doi:10.1029/2021JE006997

Ogawa, M., Yanagisawa, T. (2011) Numerical models of Martian mantle evolution induced by magmatism and solid-state convection beneath stagnant lithosphere. J Geophys Res Planets 116(E8). doi:10.1029/2010JE003777

Ogawa, M., Yanagisawa, T. (2012) Two-dimensional numerical studies on the effects of water on Martian mantle evolution induced by magmatism and solid-state mantle convection. J Geophys Res Planets 117(E6). doi:10.1029/2012JE004054

Ogawa, M., Yanagisawa, T. (2014). Mantle evolution in Venus due to magmatism and phase transitions: From punctuated layered convection to whole‐mantle convection. J Geophys Res Planets 119(4):867-883. Doi: 10.1002/2013JE004593

Ribe, N.M. (1985) The deformation and compaction of partial molten zones. Geophys J Int 83(2):487–501. doi:10.1111/j.1365-246X.1985.tb06499.x

Richter, F.M., Daly, S.F. (1989) Dynamical and chemical effects of melting a heterogeneous source. J Geophys Res Solid Earth 94(B9):12499–12510. doi:10.1029/JB094iB09p12499

Sakamaki, T., Ohtani, E., Urakawa, S., Suzuki, A., Katayama, Y. (2010) Density of dry peridotite magma at high pressure using an X-ray absorption method. Am Mineral 95(1):144–147. doi:10.2138/am.2010.3143

Scott, D.R., Stevenson, D.J. (1986) Magma ascent by porous flow. J Geophys Res Solid Earth 91(B9):9283–9296. doi:10.1029/JB091iB09p09283

Scott, D.R., Stevenson, D.J. (1989) A self-consistent model of melting, magma migration and buoyancy-driven circulation beneath mid-ocean ridges. J Geophys Res Solid Earth 94(B3):2973–2988. doi:10.1029/JB094iB03p02973

Sobolev, S.V., Sobolev, A.V., Kuzmin, D.V., Krivolutskaya, N.A., Petrunin, A.G., Arndt, N.T., Vasiliev, Y.R. (2011) Linking mantle plumes, large igneous provinces and environmental catastrophes. Nature 477(7364):312–316. doi:10.1038/nature10385

Spiegelman, M. (1993a) Flow in deformable porous media. Part 1 Simple analysis. J Fluid Mech 247:17–38. doi:10.1017/S0022112093000369

Spiegelman, M. (1993b) Flow in deformable porous media. Part 2 numerical analysis–the relationship between shock waves and solitary waves. J Fluid Mech 247:39–63. doi:10.1017/S0022112093000370

Spiegelman, M., Kelemen, P.B., Aharonov, E. (2001) Causes and consequences of flow organization during melt transport: The reaction infiltration instability in compactible media. J Geophys Res Solid Earth 106(B2):2061–2077. doi:10.1029/2000JB900240

Spiegelman, M., McKenzie, D. (1987) Simple 2-D models for melt extraction at mid-ocean ridges and island arcs. Earth Planet Sci Lett 83(1–4):137–152. doi:10.1016/0012-821X(87)90057-4

Steinberger, B., Antretter, M. (2006) Conduit diameter and buoyant rising speed of mantle plumes: Implications for the motion of hot spots and shape of plume conduits. Geochem Geophys Geosyst 7(11). doi:10.1029/2006GC001409

Stevenson, D.J. (1989) Spontaneous small-scale melt segregation in partial melts undergoing deformation. Geophys Res Lett 16(9):1067–1070. doi:10.1029/GL016i009p01067

Suzuki, A., Ohtani, E. (2003) Density of peridotite melts at high pressure. Phys Chem Miner 30(8):449–456. doi:10.1007/s00269-003-0322-6

Tackley, P.J., Stevenson, D.J. (1993) A Mechanism for Spontaneous Self-Perpetuating Volcanism on the Terrestrial Planets. In: Stone, D.B., Runcorn, S.K. (eds) Flow and Creep in the Solar System: Observations, Modeling and Theory. NATO ASI Ser 391. Springer, Dordrecht. doi:10.1007/978-94-015-8206-3_19

Taito, H., Kameyama, M., Miyagoshi, T., Ogawa, M. (2025) Two-dimensional numerical experiments on mantle convection with stress-history-dependent rheology: Toward self-consistent reproduction of plate tectonics. Earth Planets Space *in press*.

Turcotte, D.L., Schubert, G. (2014) Geodynamics. Cambridge University Press. doi: 10.1017/CBO9780511843877

U, K., Hasumi, H., Ogawa, M. (2022) Effects of magma-generation and migration on the expansion and contraction history of the Moon. Earth Planets Space 74(78):78. doi:10.1186/s40623-022-01631-4

U, K., Kameyama, M., Ogawa, M. (2023) The volcanic and radial expansion/contraction history of the Moon simulated by numerical models of magmatism in the convective mantle. J Geophys Res Planets 128:e2023JE007845. doi:10.1029/2023JE007845

U, K., Kameyama, M., Nishiyama, G., Miyagoshi, T., Ogawa, M. (2025) Long-Lasting Volcanism of the Moon Aided by the Switch in Dominant Mechanisms of Magma Ascent: Role of Localized Radioactive Enrichment in a Numerical Model of Magmatism and Mantle Convection. Geophys Res Lett 52(8). doi:10.1029/2025GL115215

Wolfe, C.J., Bjarnason, I.Th., VanDecar, J.C., Solomon, S.C. (1997) Seismic structure of the Iceland mantle plume. Nature 385(6613):245–247. doi:10.1038/385245a0

Xu, Y., Lin, Y., Wu, P., Namur, O., Zhang, Y., Charlier, B. (2024) A diamond-bearing core-mantle boundary on Mercury. Nat Commun 15(1):5061. doi:10.1038/s41467-024-49305-x

Yanagisawa, T., Kameyama, M., Ogawa, M. (2016) Numerical studies on convective stability and flow pattern in three-dimensional spherical mantle of terrestrial planets. Geophys J Int 206(3):1526–1538. doi:10.1093/gji/ggw226

Yoshizaki, T., McDonough, W.F. (2020) The composition of Mars. Geochim Cosmochim Acta 273:137–162. doi:10.1016/j.gca.2020.01.011